%% file: main.tex
\documentclass[runningheads]{llncs}
\usepackage[T1]{fontenc}
\usepackage{tikz}
\usetikzlibrary{arrows,positioning}
\usepackage{graphicx}
\usepackage{fancyvrb}
\usepackage{xcolor}
\usepackage{booktabs}
\usepackage{multirow}
\usepackage{adjustbox}
\usepackage{listings}
\usepackage{wrapfig}

\lstdefinestyle{customc}{
  belowcaptionskip=1\baselineskip,
  breaklines=true,
  language=C,
  showstringspaces=false,
  basicstyle=\footnotesize\ttfamily,
  keywordstyle=\bfseries\color{green!40!black},
  commentstyle=\color{gray!40!black},
  identifierstyle=\color{blue},
  stringstyle=\color{orange},
  numbers=left,                   
  numbersep=5pt,                   
  numberstyle=\tiny\color{gray},
}

\usepackage{hyperref}
\newcommand{\orc}{\includegraphics[height=\fontcharht\font`A]{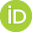}}
\newcommand{\orcid}[1]{\href{https://orcid.org/#1}{\orc}}

\begin{document}
\title{Can ChatGPT support software verification?}

\author{Christian Janßen\inst{1}  \and
Cedric Richter\inst{1}\orcid{0000-0003-2906-6508} \and
Heike Wehrheim\inst{1}\orcid{0000-0002-2385-7512}}

\authorrunning{C.~Janßen et al.}
\institute{Carl-von-Ossietzky Universität Oldenburg, Germany  \\ 
\email{\{christian.janßen1, cedric.richter, heike.wehrheim\}@uol.de} }

\maketitle              
\begin{abstract}
Large language models have become increasingly effective in software engineering
tasks such as code generation, debugging and repair. Language models 
like ChatGPT can not only {\em generate} code, but also {\em explain} its inner workings and 
in particular its correctness. This raises the question whether we can utilize 
ChatGPT to support {\em formal software verification}. 

In this paper, we take some first steps towards answering this question. 
More specifically, we investigate whether ChatGPT can generate {\em loop invariants}. 
Loop invariant generation is a core task in software verification, and 
the generation of {\em valid} and {\em useful} invariants would likely help formal verifiers. 
To provide some first evidence on this hypothesis, 
 we ask ChatGPT 
to annotate 106 C programs with  loop invariants. We check
validity and usefulness of the generated invariants by passing them to two verifiers, 
\textsc{Frama-C} and CPAchecker. Our evaluation shows that ChatGPT is able to produce
valid and useful invariants allowing \textsc{Frama-C} to verify tasks that it could not
solve before. Based on our initial insights, we propose ways of combining ChatGPT (or 
large language models in general) and software verifiers, and discuss current limitations and open issues. 

\end{abstract}

\input{chapters/intro}

\input{chapters/approach}

\input{chapters/evaluation}
\input{chapters/limitations}
\input{chapters/conclusion}

 \bibliographystyle{splncs04}
 \bibliography{reference}

\appendix
\input{chapters/appendix}

\end{document}

%% file: chapters/intro.tex
\section{Introduction}

Large language models (LLMs)~\cite{DBLP:conf/nips/BrownMRSKDNSSAA20,DBLP:journals/tmlr/WeiTBRZBYBZMCHVLDF22,DBLP:conf/nips/Ouyang0JAWMZASR22} are increasingly employed to support software engineers in the generation, testing and repair of code~\cite{DBLP:journals/corr/abs-2107-03374,DBLP:conf/iclr/ChenZNZLLC23,DBLP:conf/icse/JiangLLT23}. 
Generative AI can, however, not only generate code, but also provide explanations of the inner workings of code and 
give arguments about its correctness. This raises the question whether LLMs can also support {\em formal software verification}. 

In this paper, we provide a first step towards answering this question. In general, one can imagine various ways of supporting verifiers, 
depending on the verification approach they employ. Central to all verifiers are, however, techniques for dealing with loops. 
Specifically, for abstracting the behaviour of loops verifiers aim at computing {\em loop invariants}. 
Our first step in evaluating ChatGPT's usefulness for software verification  is thus the generation of loop invariants. 

To this end, we ask ChatGPT to annotate C-programs with loop invariants. We have chosen 106 C-programs from the {\sf{Loops}} category 
of the annual competition on software verification~\cite{DBLP:conf/tacas/Beyer23}. To enable the usage of these invariants by tools (verifiers), 
we needed  the invariants to be given in some formal language. For this, we have chosen ACSL~\cite{acsl_reference},  
a design-by-contract like annotation language for C. 
Initial experiments confirmed that ChatGPT ``knows'' ACSL. 
The main part of our experiments then concerned the evaluation of the invariants with respect to (a) {\em validity} and (b) {\em usefulness} 
for verifiers. 
The first aspect required checking whether a proposed invariant is actually a proper invariant, i.e., whether the computed predicate 
holds at the beginning of the loop and after every loop iteration. 
We employ the state-of-the-art interactive verifier \textsc{Frama-C}~\cite{DBLP:journals/cacm/BaudinBBCKKMPPS21} for this validity checking.
For evaluating the usefulness of invariants, we provided two state-of-the-art verifiers (\textsc{Frama-C} SV~\cite{DBLP:conf/tacas/BeyerS22} and CPAchecker~\cite{DBLP:conf/cav/BeyerK11}) with the code annotated by the proposed invariant,  
and evaluated whether the verifiers can then solve verification tasks which they could not solve without the invariant\footnote{In case of CPAchecker, we restrict CPAchecker's own invariant generation facilities as to be able to see the plain effect of the generated invariant.}.
Our results confirm that ChatGPT can support software verifiers by providing valid and useful loop invariants, but also show that more work needs to be done -- both conceptually and practically -- to have LLMs provide a significant support for software verification.

%% file: chapters/approach.tex
\section{Invariant Generation with Large Language Models}
Our goal is to provide initial insights into the capabilities of large language models, specifically ChatGPT,
to support formal software verification. For this, we propose the task of loop invariant generation.

\begin{figure}[t]
\texttt{Prompt>} Compute a loop invariant for the following program! 

\begin{adjustbox}{width=.4185\textwidth}
\begin{lstlisting}[language=C, style=customc]
void func(unsigned int n)
{
  unsigned int x=n, y=0;
  //@ loop invariant [mask];
  while(x>0) {
    x--; y++;
  }
  assert(y==n);
}
\end{lstlisting}
\end{adjustbox}

Infilling provided by ChatGPT: \texttt{x+y==n} 
\caption{Example task: \texttt{loops/count\_up\_down-1}. 
} 
\label{fig:ex} 
\vspace{-1.5em}
\end{figure} 

\smallskip 
\noindent\textbf{Loop invariant generation.} 
The goal of loop invariant generation is to generate  
{\em valid} and {\em useful} loop invariants for a given program.
A valid loop invariant is an invariant that (1) holds
true before the first loop execution and (2) after each
loop iteration. A useful loop invariant is a valid
loop invariant that is also useful for proving the given program correct.

To understand this, let us consider the example task 
shown in Figure~\ref{fig:ex}. Here, the large language model is tasked 
to analyze the given program and to propose a   loop invariant.
For the given program, the invariant \texttt{x + y == n} represents
a {\em valid} loop invariant:  As \texttt{x} is initialized to \texttt{n}
and \texttt{y}   to \texttt{0}, the invariant holds true (1) before
the first loop execution. The invariant furthermore holds true (2) after each
loop iteration as \texttt{y} is incremented each time \texttt{x} is decremented.

The provided loop invariant also is a {\em useful} loop invariant: As \texttt{x == 0} at the
end of the loop execution and \texttt{x + y == n} holds true after
the loop execution, we can deduce that the assertion \texttt{y == n} is not violated after the loop execution.
The invariants \texttt{x <= n} and \texttt{y >= 0}
also represent valid loop invariants but they are not useful for proving the program correct. 

\smallskip
\noindent The idea is now to let ChatGPT  generate such loop invariants. 
To this end, we need to tell ChatGPT what its task is. 
As briefly mentioned in the introduction, we 
expect ChatGPT to give loop invariants in the form of ACSL (ANSI C Specification Language~\cite{acsl_reference}) assertions.
ACSL is  a specification language for C and offers a number of keywords for specifications in a 
design-by-contract style. Among others, there is the keyword \texttt{loop invariant}. 
ACSL specifications are written inside comments of the form \verb+//@+. 
Besides the plain code, Figure~\ref{fig:ex} also shows the {\em prompt} used 
to tell ChatGPT its task (first line), and the code location and form of the invariant we expect to be generated (\verb+//@ loop invariant [mask]+)\footnote{Prompt and answer from ChatGPT are abbreviated to fit the figure; the full prompt is given in the appendix.}. 
We thus phrase the task as an {\em infilling problem}~\cite{DBLP:conf/iclr/FriedAL0WSZYZL23}, i.e., we require ChatGPT to fill in some meaningful contents for \verb+[mask]+. 
In this example, ChatGPT returns the above discussed invariant. 
We arrived at this form of stating the task after several experiments with different prompts. 
 
\smallskip 
\noindent\textbf{Feeding loop invariants into verifiers.} 
For evaluation of the generated invariants, we need to determine their validity and usefulness.
To this end, we first of all need to feed them into some verifier.
Interactive verifiers natively provide ways of feeding in such inputs. 
In an interactive verification run, a software engineer provides program annotations (e.g., invariants) and the verifier tries to 
prove that some  given specifications are never violated\footnote{There exists a variety of properties that can be checked via verification; we focus here on checking for violations of assertions.}.

In this work, our goal is to evaluate the ability of large language models 
to support verifiers. Therefore, we replace the software engineer
by ChatGPT and let it interact with the interactive verifier.
Currently, the language model only interacts by exchanging loop invariants
(which is inline with our evaluation goal). However, in future work it could
be interesting to let the language model generate other types of annotations. 

During our evaluation, we use the interactive verifier \textsc{Frama-C}~\cite{DBLP:journals/cacm/BaudinBBCKKMPPS21}
to evaluate the validity and usefulness of the provided invariants. 
For evaluating the usefulness, we furthermore employ an automatic
verifier (CPAchecker~\cite{DBLP:conf/cav/BeyerK11}). 
To also allow for interaction in this case, we employ ACSL2Witness~\cite{DBLP:conf/sefm/BeyerSU22} to convert the ACSL annotated program to a correctness witness which CPAchecker is then able to use in its verification. 

\smallskip 
\noindent\textbf{Related work.} There are only a few works
that address invariant generation via machine learning.
The work in \cite{DBLP:conf/icml/PeiBSSY23} uses large language models 
to predict invariants of Java programs. They specifically trained large
language models to predict Daikon~\cite{DBLP:journals/scp/ErnstPGMPTX07} 
generated invariants. Their evaluation does not consider validity or usefulness
of the generated invariants but only concerns whether Daikon invariants can 
be recovered. In contrast in this work, we rely on instruction-tuned large language models
such as ChatGPT {\em without} any training and we use formal verification approaches to evaluate
the validity and usefulness of loop invariants generated for C code.

Many approaches~\cite{DBLP:conf/cav/SiNDNS20,DBLP:conf/pldi/PadhiSM16,DBLP:conf/popl/0001NMR16,DBLP:conf/nips/SiDRNS18}, which are related to or based on Syntax-Guided Synthesis, have addressed invariant generation via machine learning techniques. However, most of
the existing techniques rely on traditional machine learning or graph neural network
based techniques instead of large language models. We are interested in the capabilities
of large language models in supporting software verifiers. 

Beyond invariants, there also exist other ways to support software verifiers. For example, the work in \cite{DBLP:conf/sigsoft/AlonD22,DBLP:conf/sigsoft/GiacobbeKP22} supports verifiers with neural-network based termination analyses. However, these approaches are often deeply
integrated. We chose loop invariant generation as many software verifiers already
support the exchange of invariants.

%% file: chapters/evaluation.tex
\section{Evaluation}
We evaluate  ChatGPT on the task of loop invariant generation in C code. For the evaluation, we use
a benchmark of 106 verification tasks taken from the SV-COMP {\sf{Loops}} category~\cite{DBLP:conf/tacas/Beyer23}. 
We have chosen all tasks which (a)  have ACSL annotations (to be able to compare the generated with manually constructed invariants), (b) have  
one loop only and (c) are correct, i.e., the assertions in the code are valid.  
During our evaluation, we remove all ACSL invariant annotations and let ChatGPT regenerate them. 
Now, based on our evaluation setup we aim to answer the following research question:
\begin{center}
Can ChatGPT support software verifiers with valid and useful loop invariants?
\end{center}

\smallskip 
\noindent\textbf{Experimental setup.} For generating loop invariants, we employ
the ChatGPT (GPT-3.5) snapshot from June 2023.  The model is queried via the OpenAI API\footnote{https://platform.openai.com/, accessed in Sept. 2023}. During our evaluation,
we set the sampling temperature\footnote{The temperature controls the randomness of ChatGPT's outputs; a lower temperature leads to more deterministic outputs.} of ChatGPT to $0.2$ and sample up to $k$ ($k = 5$) completions per task.
We collect all invariants by parsing the generated completions with the infillings. 

For checking the validity of the generated invariants, we use the interactive verifier \textsc{Frama-C}~\cite{DBLP:journals/cacm/BaudinBBCKKMPPS21}. 
We annotate each task with one of the $n$ generated invariants. In total, we thus generate up to $n$ annotated 
versions of each tasks which we use for validation. 
We count loop invariants as validated only if \textsc{Frama-C} WP can validate them within 10s\footnote{
Note that a negative answer of \textsc{Frama-C} does not necessarily mean that the candidate invariant is invalid.}. 

For evaluating the usefulness of the generated invariants, we now annotate the task with the validated
invariants from the previous step.
If multiple invariants are validated per task, we conjunct them to a single invariant and annotate
the task with the conjuncted invariant\footnote{The logical conjunction of two valid invariants is again a valid invariant.}. 
As verifiers, we consider the interactive verifier \textsc{Frama-C} SV~\cite{DBLP:conf/tacas/BeyerS22}\footnote{ 
\textsc{Frama-C} SV is a version of \textsc{Frama-C} specifically configured to work well
on SV-COMP task.} and the automatic verifier CPAchecker~\cite{DBLP:conf/cav/BeyerK11}. 
We configure CPAchecker to run $k$-induction without loop unrolling (similar to~\cite{DBLP:conf/sefm/BeyerSU22} to be able to see the effect of the generated invariant). Note that this restricts CPAcheckers facilities for verification. 
Finally, all verifier and validation runs are executed via BenchExec~\cite{DBLP:conf/tacas/Beyer16} on a 24-core machine with 128GB RAM running Ubuntu 22.04 with a maximum timelimit of 900s. 
\begin{table*}[t]
  \caption{Results for 106 verification tasks, divided by subcategory of the {\sf{Loops}} category (giving total number of tasks, number of successfully validated invariants, number of verified tasks per verifier using either the generated or the human provided invariant of the benchmark, and in {\color{gray} gray} the number of useful invariants) }
  \label{tab:results}
  \centering
  \begin{tabular*}{\textwidth}{l ll @{\extracolsep{\fill}} rr c | rr c rr}
    \toprule
    & \multirow{2}{*}{\shortstack[c]{Subcategory}} && \multicolumn{2}{c}{Tasks} && \multicolumn{2}{c}{\bfseries \textsc{Frama-C}} && \multicolumn{2}{c}{\bfseries $k$-induction}    \\
    \cmidrule{3-5} \cmidrule{7-8} \cmidrule{10-11}
    & && total & val-invs. && GPT invs. & Human invs. && GPT invs.  & Human invs.\\
    \midrule
    & {\sf{loop-accelaration}} && 15 & 8 && 1  {\color{gray} (1)}  & 2  {\color{gray} (2)}  && 6 {\color{gray} (3)}  & 6 {\color{gray} (3)} \\
    & {\sf{loop-crafted}} && 2 & 2 && 0  {\color{gray} (0)}  & 0  {\color{gray} (0)} && 2 {\color{gray} (0)}  & 2 {\color{gray} (0)} \\
    & {\sf{loop-industry-pat.}} && 1 & 1 && 0  {\color{gray} (0)}  & 0  {\color{gray} (0)}  && 1 {\color{gray} (0)}  & 1 {\color{gray} (0)} \\
    & {\sf{loop-invariants}} && 8 & 4 && 3  {\color{gray} (3)}  & 3  {\color{gray} (3)}  && 0  {\color{gray} (0)}   & 1  {\color{gray} (1)} \\
    & {\sf{loop-invgen}} && 3 & 3 && 0  {\color{gray} (0)}  & 0  {\color{gray} (0)}  && 0 {\color{gray} (0)}& 0 {\color{gray} (0)}\\
    & {\sf{loop-lit}} && 13 & 4 && 1  {\color{gray} (1)}  & 4  {\color{gray} (4)}  && 3 {\color{gray} (2)}& 4 {\color{gray} (3)}\\
    & {\sf{loop-new}} && 7 & 4  && 1  {\color{gray} (1)}  & 1  {\color{gray} (1)}  && 0 {\color{gray} (0)}& 0 {\color{gray} (0)}\\
    & {\sf{loop-simple}} && 1 & 1 && 1  {\color{gray} (1)}  & 1  {\color{gray} (1)}  && 1 {\color{gray} (0)} & 1 {\color{gray} (0)}\\
    & {\sf{loop-zilu}} && 22 &  18 && 10  {\color{gray} (10)}  & 11  {\color{gray} (11)}  && 11 {\color{gray} (6)} & 10 {\color{gray} (5)}\\
    & {\sf{loops}} && 13 & 13  && 5  {\color{gray} (5)}  & 6  {\color{gray} (6)}  && 8 {\color{gray} (1)}  & 8 {\color{gray} (1)} \\
    & {\sf{loops-crafted-1}} && 21 & 17 &&  0  {\color{gray} (0)} & 0  {\color{gray} (0)}  && 4 {\color{gray} (3)}  & 7 {\color{gray} (6)} \\
    \midrule
    & total && 106 & 75 && 22 {\color{gray} (22)} & 28 {\color{gray} (28)} && 36 {\color{gray} (15)} & 40 {\color{gray} (19)}  \\
    \bottomrule
  \end{tabular*}
  \vspace{-2em}
\end{table*}

\smallskip 
\noindent\textbf{Results.} Our main results are shown in Table~\ref{tab:results}. On the left side of the table, we show the total number of tasks per category (total) and the number of tasks where at least one of the generated invariants can be validated (val-invs.). On the right side of the table, we report on the verification results obtained from executing \textsc{Frama-C} and CPAchecker (using $k$-induction without loop unfolding)
on the verification tasks with at least one validated invariant. We report the total number of tasks that can be verified with a ChatGPT provided invariant (GPT invs.) and a human provided invariant (Human invs.), i.e., the ACSL invariant given in the benchmark.  In addition, we also report the number of useful
invariants in {\color{gray} gray brackets}. Useful here means that the verifier cannot complete the verification task without the invariant. 

\smallskip 
\noindent{\em ChatGPT can generate valid loop invariants.} We find that ChatGPT can generate valid loop invariants
for 75 out of 106 tasks (as validated by \textsc{Frama-C}). Note that ChatGPT proposes loop invariant candidates for all 106 tasks
and by manual inspection we found that some of the generated loop invariant candidates are still {\em meaningful}, even though they are not validated by 
\textsc{Frama-C}. An example is shown in Figure~\ref{fig:ex-fail}. ChatGPT produces a meaningful loop invariant candidate, but \textsc{Frama-C} rejects the candidate due to technical reasons\footnote{\textsc{Frama-C} reports an invalid conversion from integer type to a floating point type due to the \texttt{pow} operator and thereby fails.}. The human-annotated invariant avoids this problem
by enumerating all variable assignments. In total, we found by manual inspection that 10 out of 31 invariant candidates not validated by \textsc{Frama-C} are meaningful. 

Interestingly, we found during our manual inspection that ChatGPT in many cases seems to apply a set of useful {\em heuristics} 
to determine loop invariant candidates. One of the most successful heuristic applied by ChatGPT on our benchmark is the 
{\em copy assertion} heuristic. Here, ChatGPT proposes an invariant that is equivalent to a condition
found in a nearby assertion. The heuristic is applied in 30 out of 106 tasks and 23 of the resulting invariants are validated.
\begin{figure}[t]
\begin{adjustbox}{width=.5\textwidth}
\begin{lstlisting}[language=C, style=customc]
void func() {
  unsigned int x = 0, y = 1;
  //@ loop invariant [mask];
  while (x < 6) { x++; y *= 2; }
  assert(y % 3 != 0);
}
\end{lstlisting}
\end{adjustbox}

Infilling provided by ChatGPT: \verb+x <= 6 && y == pow(2, x)+ \\ 
Human: \verb+(x==0 && y==1) || (x==1 && y==2) ||  (x==2 &&  ==4) || ... + 
\caption{Example task: \texttt{loop-accelaration/underapprox\_1-2}}
\label{fig:ex-fail} 
\vspace{-2em}
\end{figure} 

\smallskip 
\noindent{\em ChatGPT can support verifiers with useful loop invariants.} We find that ChatGPT can produce useful invariants
that can support software verifiers in their verification tasks. In comparison to the human-provided invariants, ChatGPT produced
useful invariants for 22 out of 28 tasks in the case of \textsc{Frama-C} and for 15 out of 19 tasks in the case of CPAchecker's $k$-induction.
Interestingly, we find one example in the {\sf{loop-zilu}} subcategory where the invariant proposed by ChatGPT is more useful for CPAchecker than the human annotated invariant. The example is shown in Figure~\ref{fig:ex-better}. Here, ChatGPT proposes the invariants \texttt{j >= 0} and \texttt{k >= 0} conjuncted with the human-provided invariant which is obviously useful to prove that \texttt{k >= 0} holds true at the end of the loop. Note that, while this seems to be a case where the copy assertion heuristics is effective, \textsc{Frama-C} does not validate the invariant candidate \texttt{k >= 0} alone. The conjunction with \texttt{j<=n \&\& k>=n-j} is important to validate the invariant. Still, by manual inspection we find that the copy assertion heuristic of ChatGPT is effective for providing useful invariants in 11 out of 22 cases for \textsc{Frama-C} and in 5 out of 15 cases for $k$-induction.

%% file: chapters/limitations.tex
\section{Limitations and Open Issues}
We discuss limitations and open issues in using large language models for supporting software verifiers.

\begin{figure}[t]
\begin{adjustbox}{width=.62\textwidth}
\begin{lstlisting}[language=C, style=customc]
void func(int k, int j, int n) {
  if (!(n>=1 && k>=n && j==0)) return;
  //@ loop invariant [mask];
  while (j<=n-1) { j++; k--; }
  assert(k>=0);
}
\end{lstlisting}
\end{adjustbox}

Infilling provided by ChatGPT: \verb+j >= 0 && k >= 0 && j <= n && k >= n - j+ \\ 
Human: \texttt{j <= n <= k + j}
\caption{Example task: \texttt{loop-zilu/benchmark04\_conjunctive}. }
\label{fig:ex-better} 
\vspace{-1.4em}
\end{figure} 

\smallskip
\noindent\textbf{Cooperation between Language Model and Software Verifier.} 
Our evaluation has shown that large language models such as ChatGPT are already capable of 
producing valid and useful loop invariants for our benchmark tasks. However, to be useful in practice,
there are several challenges we have to master. 
A key challenge is the communication and cooperation between large language model and software verifier.
Currently, we have implemented a top-down approach for invariant generation, i.e., we start by querying
the language model for invariant candidates, validate them and then provide them to a verifier. 
The LLM has no knowledge about the specifics of the underlying validator or the verifier used in the process.
This can ultimately hinder the large language model from generating valid (as validated by the validator) 
or useful (as determined by the verifier) loop invariants. During our evaluation, we already have 
encountered an example where this knowledge gap leads
to meaningful but not validated invariant candidates (see Figure~\ref{fig:ex-fail}). Here, the language
model has no knowledge about the specifics of the validator used (\textsc{Frama-C}) or at least
is not informed that the proposed expression leads to a parsing error.
Communicating this information allows the large language model to self-debug~\cite{DBLP:journals/corr/abs-2304-05128} its invariant proposals
and thereby propose invariant candidates that are validated by the validator and that are useful for the verifier.
For example, if we report the implicit conversion error back to ChatGPT, 
it generates a new invariant candidate (\texttt{y == 1 << x}) for our example in Figure~\ref{fig:ex-fail} 
that is validated by our validator.
\begin{wrapfigure}{r}{0.5\textwidth}
\vspace*{-15pt}
\begin{adjustbox}{max width=0.5\textwidth}
\input{figures/overview-approach}
\end{adjustbox}
\caption{Conceptual overview.}
\label{fig:overview}
\vspace*{-20pt}
\end{wrapfigure}
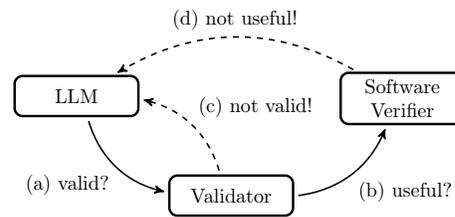 
Overall, we envision a cooperative approach between large language model, invariant validator and software verifier as shown
in Figure~\ref{fig:overview}. In an inner loop, the large language model cooperates with the validator to identify
valid loop invariants. Here, the language model proposes invariant candidates, obtains feedback from the validator
and refines its invariant suggestion. In the outer loop, the language model cooperates in the same way 
with the software verifier to find useful loop invariants. 
This work already implements (a) the validation of invariant candidates and (b) the verification with useful invariants. The key challenge
is now to determine which feedback is needed from (c) the validator or (d) the software verifier to effectively guide the language
model to valid and useful invariants. 

Finally, we can envision that our approach to language model and verifier cooperation may be useful beyond invariant generation. For example, TriCo~\cite{DBLP:conf/isola/AhrendtGJR22} proposes
to check the conformity between implementation and code specification with a verifier. A large
language model could react to conformity violations and repair either the implementation
or the specification.

\smallskip
\noindent\textbf{Unified assertion language. }
Our approach for invariant generation requires that large language models, validators and
software verifiers communicate invariants with a common specification language (e.g., ACSL in our case). However, in practice, there exists a zoo of interactive verifiers such as \textsc{Dafny}~\cite{DBLP:conf/lpar/Leino10}, \textsc{Frama-C}~\cite{DBLP:journals/cacm/BaudinBBCKKMPPS21}, \textsc{KeY}~\cite{DBLP:journals/sosym/AhrendtBBBGHMMRSS05}, \textsc{KIV}~\cite{DBLP:journals/sttt/ErnstPSHR15}, and \textsc{VeriFast}~\cite{DBLP:conf/nfm/JacobsSPVPP11} and automated software verifiers such as CBMC~\cite{DBLP:conf/tacas/ClarkeKL04}, CPAchecker~\cite{DBLP:conf/cav/BeyerK11}, Symbiotic~\cite{DBLP:journals/sttt/ChalupaSV20}, and Ultimate Automizer~\cite{DBLP:conf/cav/HeizmannHP13}. All of them implement their own custom way to communicate invariants. Therefore, we either have to find a way to unify the communication of invariants between systems or we have to define {\em transformations} that convert between communication formats. In this work, we have already employed the transformation \textsc{ACSL2Witness}~\cite{DBLP:conf/sefm/BeyerSU22} to convert ACSL to a format understandable by automated software verifiers. In the future, we plan to explore alternative transformations to support a wider range of validators and verifiers.

\smallskip
\noindent\textbf{Known limitations of LLMs.} 
Large language models have many known limitations such as hallucinations~\cite{DBLP:journals/csur/JiLFYSXIBMF23}, 
input length limitations~\cite{DBLP:conf/nips/Ouyang0JAWMZASR22}, and limited reasoning capabilities~\cite{DBLP:journals/corr/abs-2112-11446}. All of this can significantly limit 
the ability of large language models to produce valid and useful loop invariants
or to support software verifiers in general. 
However, active research is underway to overcome these limitations, and a number of proposals have 
already been made to reduce hallucinations~\cite{DBLP:journals/corr/abs-2302-12813}, increase input length~\cite{DBLP:journals/corr/abs-2306-15595}, or improve the reasoning performance~\cite{DBLP:conf/nips/Wei0SBIXCLZ22} of large language models. It would be interesting for future work to evaluate how these solutions impact the loop invariant generation abilities of large language models.

%% file: figures/overview-approach.tex
\begin{tikzpicture}[%
	node distance=1cm, auto,
    >=stealth',
    box/.style={
           rectangle,
           rounded corners,
           draw=black, very thick,
           text width=5.5em,
           minimum height=2em,
           text centered},
    pil/.style={
           ->,
           thick,
           shorten <=2pt,
           shorten >=2pt,}
]
\node[box] (llm) {LLM};
\node[box, below right=1cm and 0.5cm of llm] (val) {Validator};
\node[box, below right=-0.8cm and 3.5cm of llm] (veri) {Software\\Verifier};

\draw[pil, bend right] (llm) edge node[below left]{(a) valid?} (val);
\draw[pil, bend right] (val) edge node[below right]{(b) useful?} (veri);

\draw[pil, bend right, dashed] (val) edge node[above right]{(c) not valid!} (llm);
\draw[pil, bend right, dashed] (veri) edge node[above]{(d) not useful!} (llm);

\end{tikzpicture}

%% file: chapters/conclusion.tex
\section{Conclusion}
In this work, we provided a first step towards answering the question whether
large language models can support formal software verification. For this, we 
have evaluated ChatGPT on the task of loop invariant generation. Our evaluation
shows that ChatGPT can support software verifies by providing valid and
useful loop invariants. We plan to further improve the support for software verification
in future work by a cooperative approach that enables exchange of information
between large language models, invariant validators and software verifiers. 
In particular, we intend to develop methods for providing feedback to LLMs 
whenever candidate invariants are found to not be valid. 

%% file: chapters/appendix.tex
\section{Prompting ChatGPT}
The full prompt and the answer of ChatGPT for the example task \texttt{loops/count\_up\_down-1} is shown in Figure~\ref{fig:app-ex}. We use the same prompt for all tasks. The answer of ChatGPT can slightly vary between executions. Therefore, we generate up to $k$ answers and collect invariants from all answers via a regular expression.  
\begin{figure}[t]
\texttt{Prompt>} Compute a loop invariant for the annotated loop including [invariant] in the following C code. Please use the format of ACSL annotations and always end your response with //@ loop invariant X ; where X is the computed invariant.
\begin{lstlisting}[language=C, style=customc]
extern void abort(void);
extern void __assert_fail(const char *, const char *, unsigned int, const char *) __attribute__ ((__nothrow__ , __leaf__)) __attribute__ ((__noreturn__));
void reach_error() { __assert_fail("0", "count_up_down-1.c", 3, "reach_error"); }

void __VERIFIER_assert(int cond) {
  if (!(cond)) {
    ERROR: {reach_error();abort();}
  }
  return;
}
unsigned int __VERIFIER_nondet_uint();

int main()
{
  unsigned int n = __VERIFIER_nondet_uint();
  unsigned int x=n, y=0;
  //@ loop invariant [invariant];
  while(x>0)
  {
    x--;
    y++;
  }
  __VERIFIER_assert(y==n);
}
\end{lstlisting}
\texttt{ChatGPT>} 
\begin{lstlisting}[basicstyle=\small]
The loop invariant for the given code can be stated as follows:

```
//@ loop invariant y + x == n ;
```

Explanation:
- The loop invariant states that the sum of variables `y` and `x` is 
always equal to `n`.
- Initially, `x` is assigned the value of `n` and `y` is assigned `0`.
- In each iteration of the loop, `x` is decremented by 1 (`x--`) and 
`y` is incremented by 1 (`y++`).
- Since the loop continues until `x` becomes 0, the total number of iterations
 is equal to the initial value of `x` (which is `n`).
- Therefore, at the end of the loop, `y` will be equal to `n`.
- This loop invariant is verified by the assertion `y == n` after the loop.

Hence, the loop invariant is:

```
//@ loop invariant y + x == n ;
```
\end{lstlisting}
\caption{Full prompt and answer for example task: \texttt{loops/count\_up\_down-1}. 
} 
\label{fig:app-ex} 
\end{figure} 